\documentclass[a4paper,11pt]{article}
\usepackage{jinstpub} % for details on the use of the package, please
\usepackage[ruled,vlined]{algorithm2e}
\usepackage{wrapfig}
\usepackage{subcaption}
\usepackage{caption}
\usepackage{float}
\usepackage[mathlines]{lineno} % line numbers that work well with equations

\title{IceCube DeepCore's sensitivity to Non-Standard neutrino Interactions in the Earth}

\collaboration[c]{on behalf of the IceCube Collaboration$^*$\note[*]{Full author list and acknowledgments are available at \href{https://icecube.wisc.edu/collaboration/authors/\#collab=IceCube&date=2025-08-25&formatting=web}{icecube.wisc.edu}.}}

\author{Samyak Jain\textsuperscript{1}, Veronika Palusova\textsuperscript{2}, Thomas Ehrhardt\textsuperscript{2}, Sebastian Boser\textsuperscript{2}, and Francis Halzen\textsuperscript{1}}
\affiliation{\textsuperscript{1}University of Wisconsin - Madison,\\Madison, WI, USA}

\affiliation{\textsuperscript{2}Johannes Gutenberg University of Mainz, \\Mainz, Rhineland-Palatinate, Germany}

\emailAdd{samyak@icecube.wisc.edu}
\emailAdd{vpalusova@icecube.wisc.edu}
\emailAdd{tehrhardt@icecube.wisc.edu}
\emailAdd{sboeser@icecube.wisc.edu}
\emailAdd{halzen@icecube.wisc.edu}

\abstract{Neutrino oscillations continue to provide one of the most promising avenues for uncovering physics beyond the Standard Model. In particular, beyond-standard-model neutrino matter interactions may perturb neutrino oscillations in matter, leading to an observable signal in long baseline oscillation experiments. Moreover, such interactions can be a possible explanation of the rising tension between T2K and NOvA's $\delta_{\text{CP}}$ measurements. We examine IceCube DeepCore's sensitivity to these Non-Standard Interactions (NSI) by employing a model-independent NSI parameterization, and examine IceCube DeepCore's ability to comment on NSI being the cause of the T2K-NOvA $\delta_{\text{CP}}$ tension.}

\proceeding{Presented at the 32nd International Symposium on Lepton Photon Interactions at High Energies, Madison, Wisconsin, USA, August 25-29, 2025}

\notoc
\begin{document}
\maketitle
\flushbottom

\section{Introduction and Background}
\label{sec:introduction}
The discovery of non-zero neutrino masses guarantees the existence of new physics and/or new particles. New neutrino-matter interactions, referred to as Non-Standard Interactions (NSI), are a feature of many models that generate neutrino masses. Atmospheric neutrinos, produced via cosmic rays interactions in the atmosphere, traverse large baselines, ranging from a few kilometers up to the Earth diameter through matter, to reach IceCube. Any NSI in the Earth could influence the oscillation probabilities of atmospheric neutrinos along their baselines, making IceCube-DeepCore well suited to look for signatures of NSI.

Another indication of NSI comes from a rising tension \cite{T2K_NOvA_original} (currently $\sim 2 \sigma$ \cite{T2K_NOvA}) between the measured values of $\delta_{\text{CP}}$ in T2K \cite{T2K_result} and NOvA \cite{NOvA_result}, both of which are long-baseline oscillation experiments. However, NOvA has a much longer baseline through the Earth, making matter effects much stronger in NOvA than T2K. It has been shown that this $\delta_\text{CP}$ tension could be resolved by introducing non-zero NSI couplings $\varepsilon_{e\mu}$ and $\varepsilon_{e\tau}$ \cite{T2K_NOvA_original, T2K_NOvA}.

In this work, we will calculate IceCube-DeepCore's sensitivity to NSI with 9.28 years of data. We use a model-independent parameterization of NSI that allows for multiple non-zero NSI couplings simultaneously, and also separately calculate the sensitivities to $\varepsilon_{e\mu}$ and $\varepsilon_{e\tau}$.
\vspace{-0.2cm}
\section{Parameterizing NSI}
\label{sec:NSI_parameterization}
\vspace{-0.2cm}

In this section, we briefly describe the parameterization of NSI used in this analysis. For a more detailed discussion, see \cite{NSI_dev}. 
 One can introduce NSI in a model-independent manner via non-zero perturbative coupling strengths $\varepsilon_{\alpha\beta}$ in each term of the matter Hamiltonian \cite{elisa}
\begin{eqnarray}
H_{\text{mat}}(x) = \sqrt{2} G_F N_e(x) \begin{pmatrix}
        1+\varepsilon_{ee} &\varepsilon_{e\mu} &\varepsilon_{e\tau} \\
        \varepsilon_{e\mu}^* & \varepsilon_{\mu\mu} & \varepsilon_{\mu \tau} \\
        \varepsilon_{e\tau}^* & \varepsilon_{\mu \tau}^* & \varepsilon_{\tau\tau} 
    \end{pmatrix},
\end{eqnarray}
where $N_e$ is the electron density at position $x$, and $G_F$ is Fermi's coupling constant. The 1 in the $ee$ term accounts for the MSW effect \cite{MSW} - the modification of the matter Hamiltonian due to Standard Model neutrino-matter interactions. The coupling strengths $\varepsilon_{\alpha\beta}$ corresponds to NSI causing oscillations in the $\alpha-\beta$ sector at first order.  We can simplify this by subtracting an unphysical global phase, chosen to be $\varepsilon_{\mu\mu}\mathbb{I}$, from $H_\text{mat}$ to obtain
        \begin{eqnarray}
            H_{\text{mat}} = \sqrt{2} G_F N_e(x) \begin{pmatrix}
        1+\varepsilon_{ee} - \varepsilon_{\mu\mu}&\varepsilon_{e\mu} &\varepsilon_{e\tau} \\
        \varepsilon_{e\mu}^* & 0 & \varepsilon_{\mu \tau} \\
        \varepsilon_{e\tau}^* & \varepsilon_{\mu \tau}^* & \varepsilon_{\tau\tau} - \varepsilon_{\mu\mu} \label{Hmat_final}
    \end{pmatrix}.
    \label{std_nsi}
        \end{eqnarray}
This parameterization leaves us with 8 independent parameters - 2 real diagonal parameters, and 3 complex off-diagonal parameters (each with a magnitude and a phase). We can simplify this by decomposing $H_\text{mat}$ into rotation matrices \cite{minos_nsi}
\begin{eqnarray} 
\scriptsize
H_{\text{mat}} = Q_{\text{rel}} U_{\text{mat}} D_{\text{mat}} U_{\text{mat}}^{\dagger} Q_{\text{rel}}^{\dagger}
\quad \text{with} \quad
\begin{cases}
Q_{\text{rel}} = \text{diag}\left(e^{i\alpha_1}, e^{i\alpha_2}, e^{-i\alpha_1 - i\alpha_2}\right), \\
U_{\text{mat}} = R_{12}(\varphi_{12}) R_{13}(\varphi_{13}) \Tilde{R}_{23}(\varphi_{23}, \delta_\text{NS}), \\
D_{\text{mat}} = V_{CC}(x) \, \text{diag}(\varepsilon, \varepsilon^\prime, 0).
\end{cases}
\end{eqnarray}
Here, $\varepsilon$ and $\varepsilon^\prime$ are the eigenvalues of $H_\text{mat}$, $\varphi_{12}, \varphi_{13}, \varphi_{23}$ are rotation angles, and $\alpha_1, \alpha_2, \delta_\text{NS}$ are CP violating phases. It has been shown that in the special case of $\varepsilon^\prime = 0$, NSI effects become very weak and oscillations start to mimic vacuum oscillations \cite{minos_nsi, dragon}. We can thus place the most conservative NSI constraints in this regime. 

Upon setting $\varepsilon^\prime = 0$, $\delta_\text{NS}$ and $\varphi_{23}$ become unphysical \cite{minos_nsi, dragon}. We further neglect $\alpha_1$ and $\alpha_2$ because IceCube has little sensitivity to CP violating NSI \cite{dragon}. We are thus left with 3 NSI parameters - $\varepsilon$, $\varphi_{12}$, and $\varphi_{13}$ - this parameterization be reffered to as the Generalized Matter Potential (GMP). 

For the absence of NSI, which we will refer to as the Standard Interactions (SI) case, we have $\varepsilon = 1, \varphi_{12} = \varphi_{13} = 0$. The parameter $\varepsilon$ is an overall scaling which controls the strength of NSI effects, while non-zero $\varphi_{12}$ and $\varphi_{13}$ introduce first-order NSI-induced oscillations in the $e-\mu$ and $e-\tau$ sectors, respectively. For the case of real NSI, the most general ranges of $\varphi_{12}$ and $\varphi_{13}$ can be shown to be $(-90^\circ, 90^\circ$) each \cite{minos_nsi}. It can also be shown that oscillation probabilities depend only on the overall sign of $\varepsilon\cdot \Delta m_{31}^2$ \cite{minos_nsi}. We thus hold $\Delta m_{31}^2$ positive, and allow $\varepsilon$ to be negative to investigate the inverted mass ordering.

\vspace{-0.2cm}
\section{Data selection}
\vspace{-0.2cm}

Each event in IceCube DeepCore can be parameterized by the neutrino energy, the zenith angle $\theta$ of the neutrino, and the neutrino flavor. Interactions of different neutrino flavours produce different event morphologies in the Antarctic ice. Neutral-current interactions of all flavors produce hadronic cascades, and, at the relevant energy scales and given the resolution of IceCube DeepCore, electron and tau neutrino charged-current interactions are also seen as cascades, while muon neutrino charged-current interactions leave a track-like signature.

This analysis uses the reconstruction algorithm described in \cite{flercnn}, which has been trained on the 150,000 preselected events from 9.28 years of DeepCore data \cite{icecube_osc_sample}, to generate event counts in different morphologies (tracks, cascades, and mixed - the event cannot reliably be categorized as either a track or cascade) from atmospheric neutrinos for different NSI hypotheses, for 12 log-spaced energy bins between 5 and 100 GeV and 8 bins in $\cos\theta$ between -1 and 0. Our dataset is $\sim 4$ times larger than the 3 year dataset used for the previous DeepCore NSI analysis \cite{dragon}, and this, along with the improvement in our understanding of DeepCore systematics since the last analysis, should provide us with much higher sensitivities.
\vspace{-0.2cm}
\section{Sensitivities}
\vspace{-0.2cm}

To compare different NSI hypotheses, we use the modified $\chi^2$ as our Test-Statistic (TS) - 
\begin{eqnarray}
    \chi^2_\text{mod} = \frac{(N_\text{obs} - N_\text{exp})^2}{\sigma^2 + N_\text{exp}},
\end{eqnarray}
where the usual $\chi^2$ is modified by the $N_\text{exp}$ term in the denominator to account for errors due to finite Monte Carlo statistics. In calculating our sensitivities, we assume the SI case to be the observed data, and calculate the TS for simulated event counts for each NSI hypothesis.

\begin{figure}[h]
\centering
\begin{subfigure}{0.4\textwidth}
  \includegraphics[width=\linewidth]{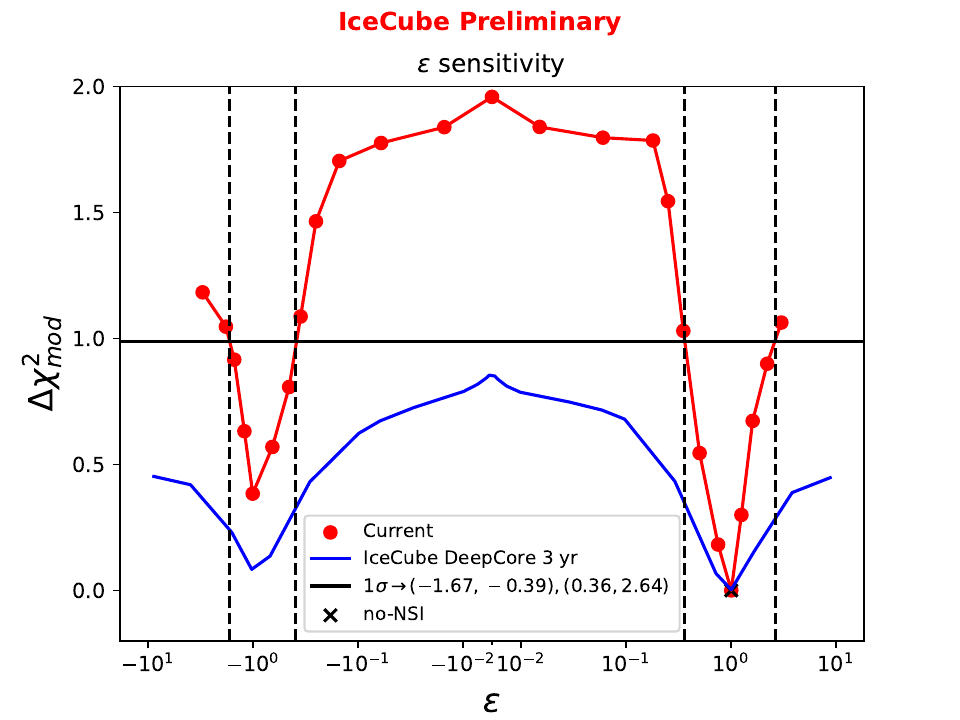}
  \caption{Sensitivity to $\varepsilon$}
  \label{epsilon_sensitivity}
\end{subfigure}%
\begin{subfigure}{0.4\textwidth}
  \includegraphics[width=\linewidth]{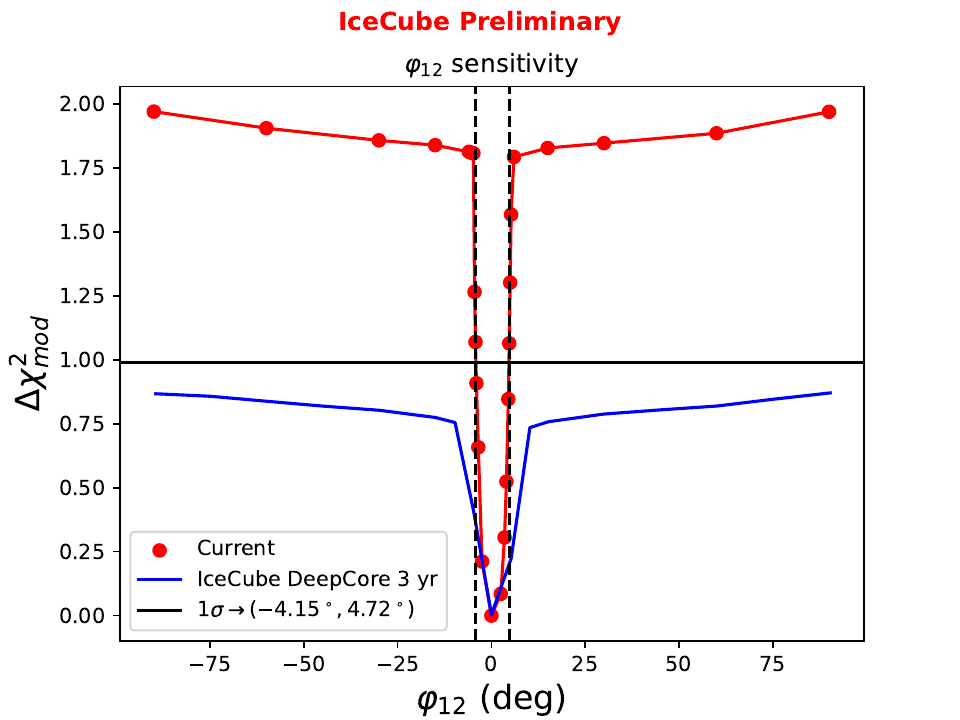}
  \caption{Sensitivity to $\varphi_{12}$}
\end{subfigure}%
\centering
\newline
\begin{subfigure}{0.4\textwidth}
  \includegraphics[width=\linewidth]{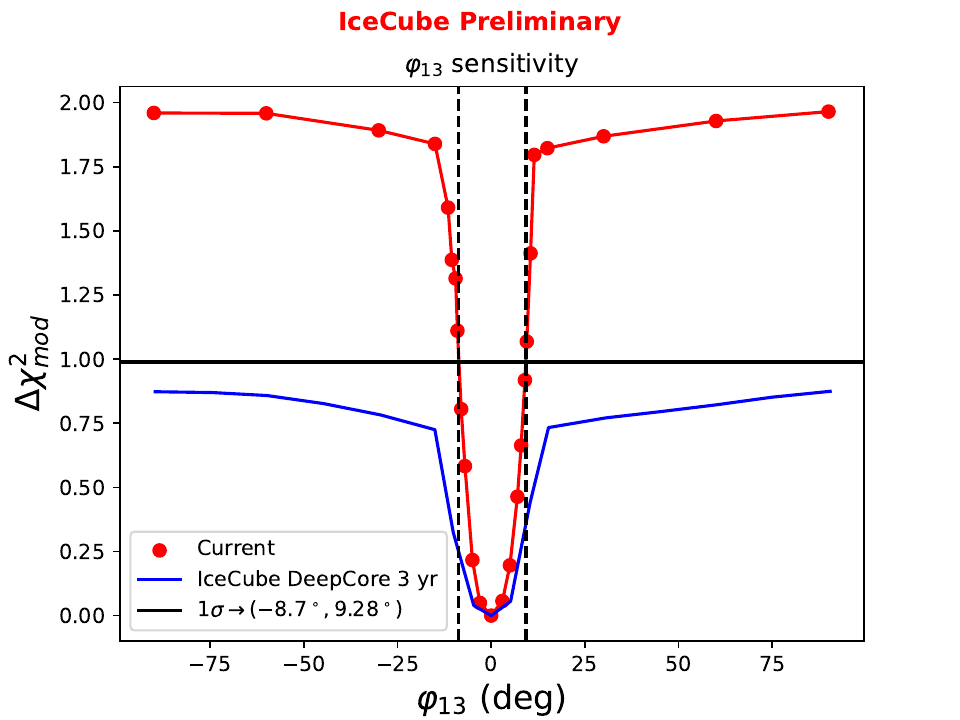}
  \caption{Sensitivity to $\varphi_{13}$}
\end{subfigure}
\caption[One parameter NSI sensitivities]{\label{sensitivities_gmp} Sensitivities to the SI hypothesis ($\varepsilon = 1, \varphi_{12} = \varphi_{13} = 0$) for each individual GMP parameter are shown in red. For comparison, the sensitivities for the previous 3 year DeepCore analysis \cite{dragon} are shown in blue.}
\end{figure}

We show our sensitivities to the Generalized Matter Potential parameters in Fig. \ref{sensitivities_gmp}. We find that our sensitivities have improved by a factor of $2-3$ over the sensitivities of the 3-year IceCube DeepCore NSI analysis \cite{dragon}, which provides the current leading constraints on the GMP parameters. In the case of SI, we would obtain $1\sigma$ constraints on $\varepsilon$ as $(-1.67, -0.39), (0.36, 2.64)$, where the first interval corresponds to the case of the inverted mass ordering. Our corresponding $1\sigma$ bounds on $
\varphi_{12}$ and $\varphi_{13}$ would be $(-4.15^\circ, 4.72^\circ)$ and $(-8.7^\circ, 9.28^\circ)$ respectively.

\begin{figure}[h!]
\centering
\begin{subfigure}{0.42\textwidth}
\includegraphics[width = \linewidth]{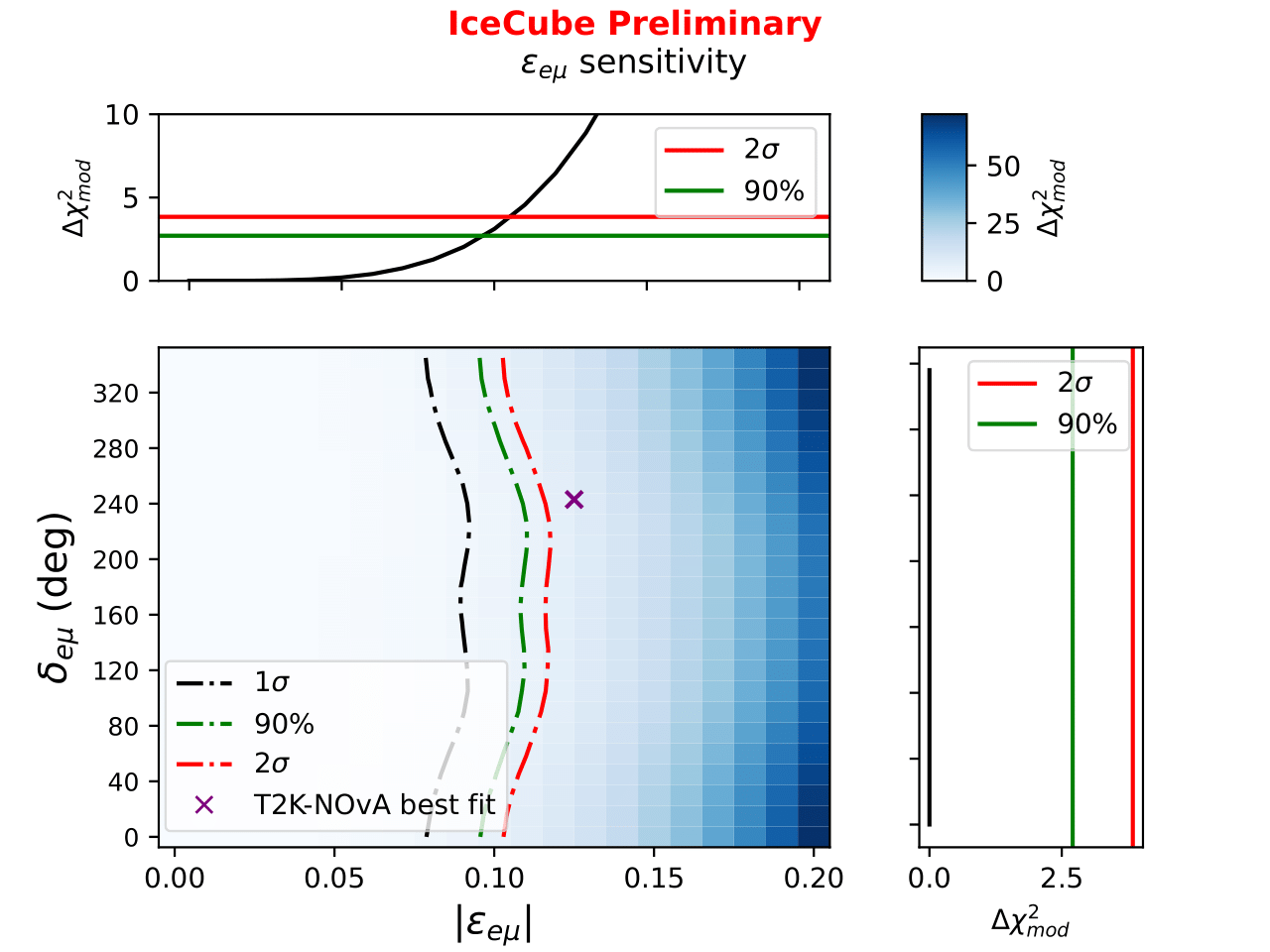}

  \subcaption{Sensitivity to $\varepsilon_{e\mu}$ \label{emu_sens}}
  
\end{subfigure} 
\begin{subfigure}{0.42\textwidth}
  \includegraphics[width = \linewidth]{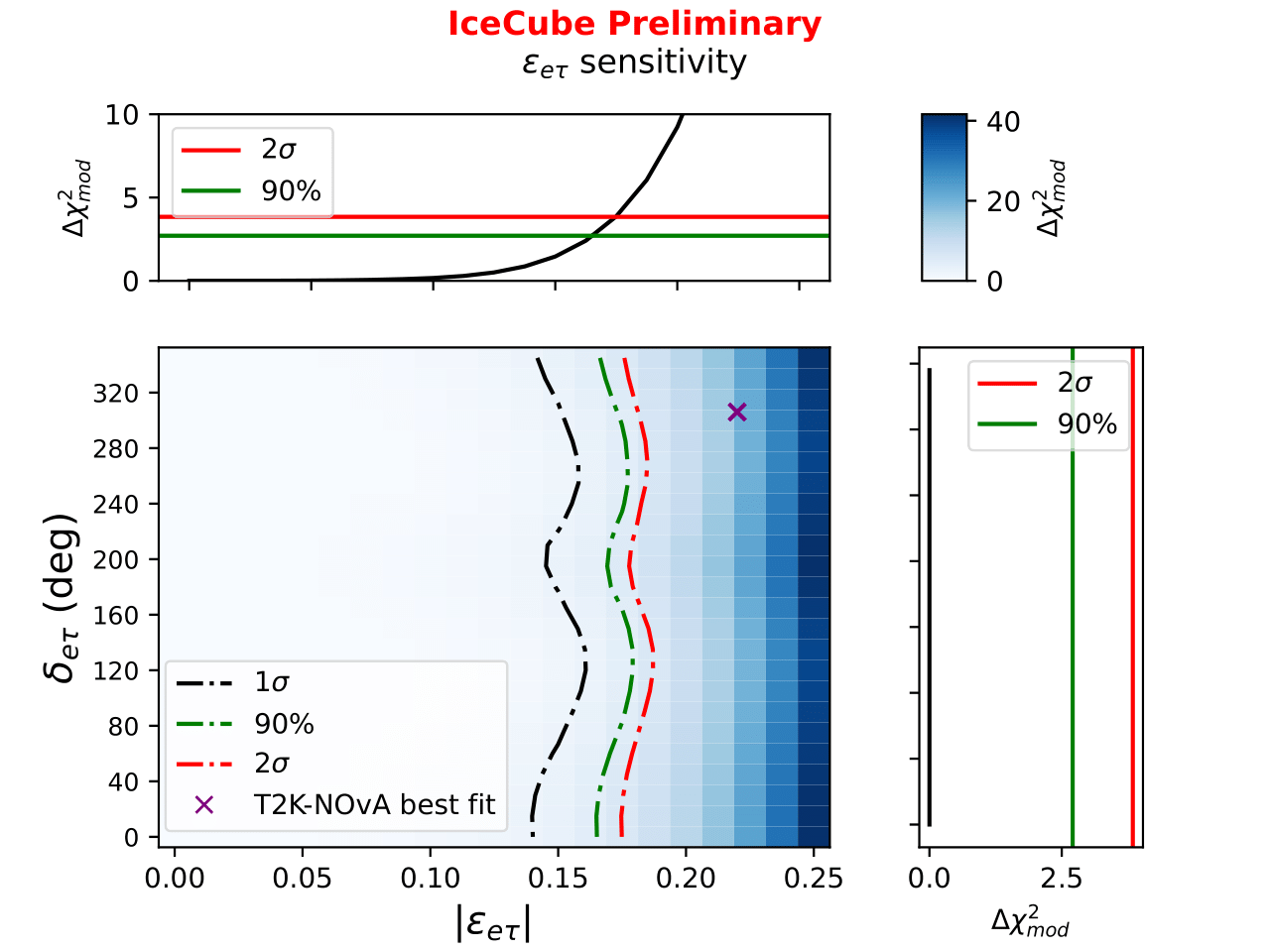}
  \subcaption{Sensitivity to $\varepsilon_{e\tau}$ \label{etau_sens}}
\end{subfigure}
\caption{Our sensitivities to $\varepsilon_{e\mu}$ and $\varepsilon_{e\tau}$ are shown in panels (a) and (b) respectively. We indicate the best-fit value of each parameter for resolving the T2K-NOvA tension with an x. The projections of the TS profile on to the magnitude and phase of each NSI parameter are shown as well.}
\label{emu_etau_sensitivities}
\end{figure}

We then consider $\varepsilon_{e\mu}$ and $\varepsilon_{e\tau}$ from the original NSI parameterization (Eq. \ref{std_nsi}), which could resolve the T2K-NOvA tension. Our sensitivities to $\varepsilon_{e\mu}$ and $\varepsilon_{e\tau}$ are shown in Fig. \ref{emu_etau_sensitivities}. We show our sensitivities as a function of the magnitude and phase of each NSI parameter, along with the projections on to the magnitude and phase separately. We find that assuming the SI case, we can rule out $\varepsilon_{e\mu}$ and $\varepsilon_{e\tau}$ best-fit values for resolving the T2K-NOvA tension (obtained from \cite{T2K_NOvA}, indicated by an x in Fig. \ref{emu_etau_sensitivities}) with a $\chi^2_\text{mod}$ of 6.81 and 20.62 respectively, which correspond to significances of $2.13\sigma$ and $4.15\sigma$ respectively with 2 degrees of freedom. In the case of SI, our $2\sigma$ bounds on the magnitude of $\varepsilon_{e\mu}$ and $\varepsilon_{e\tau}$ would be (0, 0.11) and (0, 0.175) respectively. We do not have any sensitivity to the phases because they are irrelevant for the the case of SI.

\vspace{-0.3cm}
\vspace{-0.2cm}
\section{Conclusions}
\label{sec:conc}
We present the sensitivities of IceCube DeepCore to various NSI parameters with 9.28 years of data. We expect to put leading constraints on all NSI parameters considered - for the case of the Generalized Matter Potential, our analysis is $\sim 2-3$ times more sensitive than the previous 3-year analysis \cite{dragon}. In the absence of NSI, our $1\sigma$ constraints on $\varepsilon$ would be (0.36, 2.64) and (-1.67, -0.39) for the normal and inverted mass orderings respectively, while our $1\sigma$ constraints on $
\varphi_{12}$ and $\varphi_{13}$ would be $(-4.15^\circ, 4.72^\circ)$ and $(-8.7^\circ, 9.28^\circ)$ respectively. For the case of $\varepsilon_{e\mu}$ and $\varepsilon_{e\tau}$, in the absence of NSI, our $2\sigma$ constraints on their magnitudes would be (0,0.11) and (0,0.175) respectively. With regards to the T2K-NOvA $\delta_\text{CP}$ tension, in the absence of NSI, we would be able to rule out the best-fit values from the combined T2K-NOvA fit \cite{T2K_NOvA} at $2.13\sigma$ and $4.15\sigma$ respectively. The constraints from our analysis should allow us to make strong statements about NSI being the cause of the T2K-NOvA $\delta_\text{CP}$ tension.
\vspace{-0.2cm}
% We suggest to always provide author, title and journal data:
% in short all the informations that clearly identify a document.
\bibliographystyle{JHEP}
\bibliography{refs}

\end{document}